\documentclass[fleqn,12pt,twoside]{article}
\usepackage{espcrc1}

\usepackage{graphicx}

\title{The horn, the hadron mass spectrum and the QCD phase diagram - the statistical 
model of hadron production in central nucleus-nucleus collisions}

\author{A. Andronic\address[GSI]{GSI Helmholtzzentrum f\"ur Schwerionenforschung,
D-64291 Darmstadt, Germany},
        P. Braun-Munzinger\addressmark$^,$\address[EMMI]{ExtreMe Matter Institute EMMI, D-64291 Darmstadt, Germany}$^,$\address[TU]{Technical University Darmstadt, D-64289 Darmstadt, Germany}$^,$\address{FIAS, J.W Goethe University, D-60438 Frankfurt, Germany},
        and
        J. Stachel\address{Physikalisches Institut der Universit\"at Heidelberg,
D-69120 Heidelberg, Germany}}

\begin{document}

\maketitle

One of the major goals of ultrarelativistic nuclear collision studies is to
obtain information on the QCD phase diagram \cite{pbm_wambach}.  A promising
approach is the investigation of hadron production.  Hadron yields measured in
central heavy ion collisions from AGS up to RHIC energies can be described very
well (see \cite{aa08} and refs. therein) within a hadro-chemical equilibrium 
model.  In our approach the only parameters are the chemical freeze-out 
temperature $T$, the baryo-chemical potential $\mu_b$ and the fireball volume $V$
(for a review see \cite{review}).

The main result of these investigations was that the extracted temperature
values rise rather sharply from low energies on towards $\sqrt{s_{NN}}\simeq$10 GeV
and reach afterwards constant values near $T$=160 MeV, while the baryochemical
potential  decreases smoothly as a function of energy.
The limiting temperature \cite{hagedorn85} behavior suggests a connection to 
the phase boundary and it was, indeed, argued \cite{wetterich} that the 
quark-hadron phase transition drives the equilibration dynamically, at least 
for  SPS energies and above. 
Considering also the results obtained for elementary collisions, where similar
analyses of hadron multiplicities, albeit with several additional, non-statistical 
parameters (see \cite{aa08,becattini08} and refs. therein),  yield 
also temperature values in the range of 160 MeV, alternative interpretations 
were put forward. These include conjectures  that the thermodynamical 
state is not reached by dynamical equilibration among constituents but rather 
is a generic fingerprint of hadronization \cite{stock,heinz}, or is a feature 
of the excited QCD vacuum \cite{castorina}. 

While in general all hadron yields are described rather quantitatively, a
notable exception was up-to-now the energy dependence of the $K^+/\pi^+$ ratio 
which exhibits a rather marked maximum, ``the horn'' \cite{gaz}, 
near $\sqrt{s_{NN}}\simeq$ 10 GeV \cite{na49pi}. 
Predicted first within a model of quark-gluon plasma (QGP) formation \cite{gaz},
the existence of such a maximum was also predicted \cite{pbm4} within 
the framework of the statistical model, but the observed rather sharp structure 
could not be reproduced \cite{aa05} (see also the discussion in \cite{na49pi}). 
As a consequence, the horn structure is taken in
\cite{na49pi} as experimental evidence for the onset of deconfinement and
QGP formation, and as support for the predictions of \cite{gaz}.
We have recently shown \cite{aat2} that, employing an improved hadronic mass 
spectrum, in which the $\sigma$ meson and many higher-lying resonances 
are included, leads to a sharpening the structure in the $K^+/\pi^+$ ratio, 
as will be shown below.

\begin{figure}[hbt]
\vspace{-.6cm}
\begin{tabular}{lr} \begin{minipage}{.49\textwidth}
\hspace{-.5cm}\includegraphics[width=1.\textwidth]{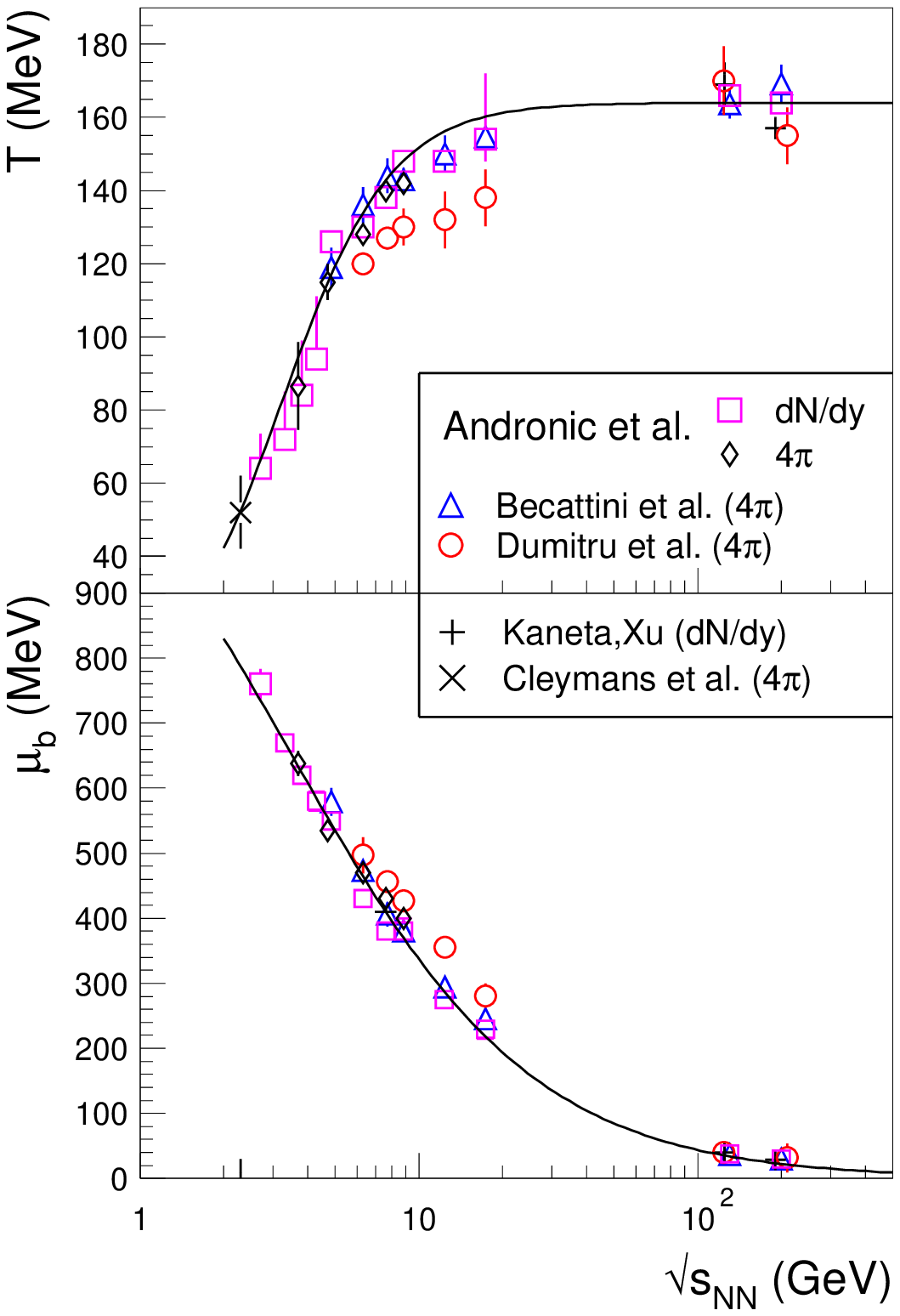}
\vspace{-.6cm}
\caption{The energy dependence of the temperature and baryon chemical potential 
at chemical freeze-out. The  results of the new fits \cite{aat2} are compared 
to the values obtained in our earlier study \cite{aa05}.
The lines are parametrizations for $T$ and $\mu_b$ (see text).}
\label{fig_tmu}
\end{minipage} &\begin{minipage}{.49\textwidth}
\vspace{-.3cm}
\hspace{-.5cm}\includegraphics[width=1.07\textwidth]{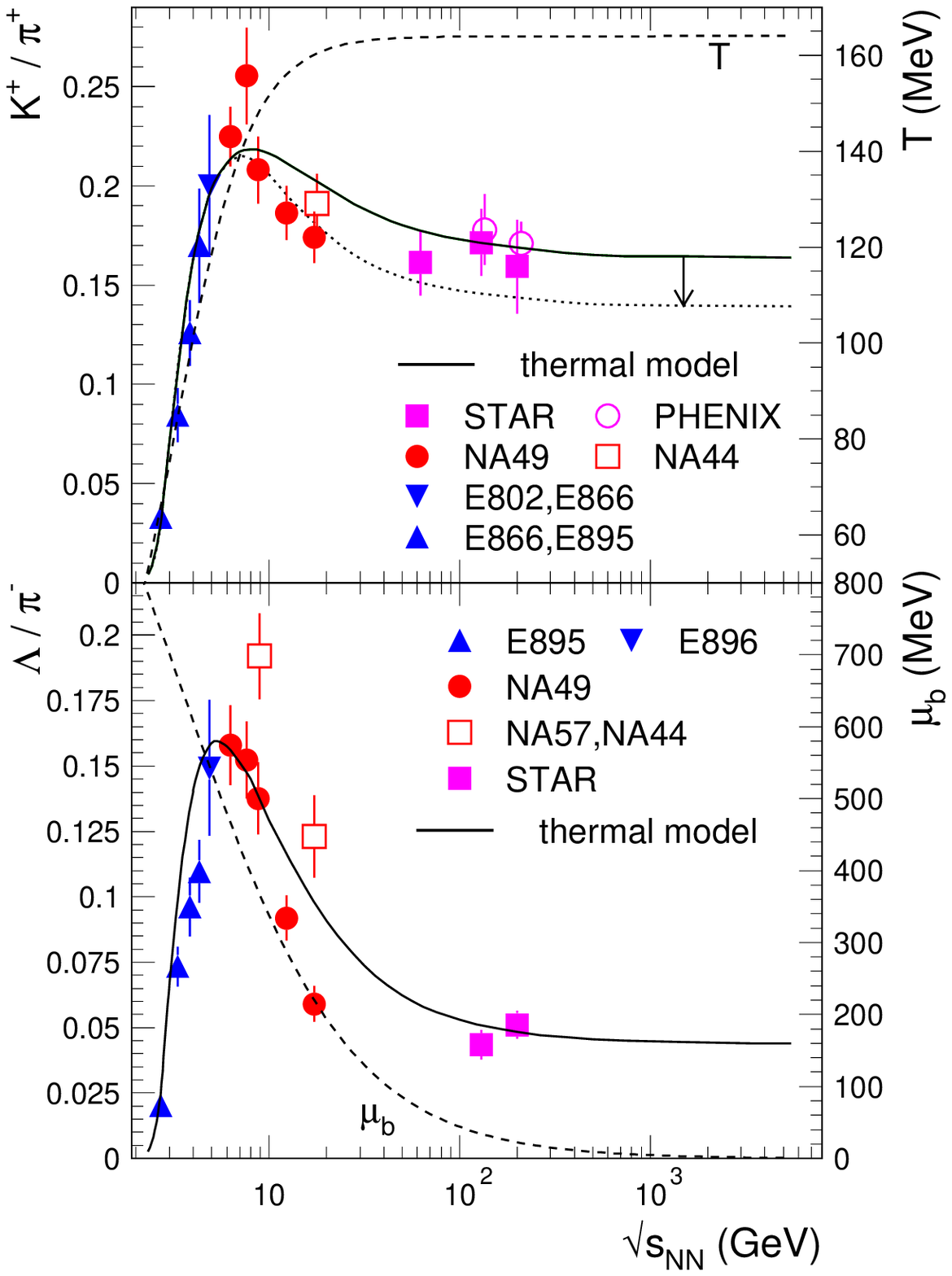}
\vspace{-.9cm}
\caption{Energy dependence of the relative production ratios $K^+/\pi^+$ 
and $\Lambda/\pi^-$. 
With the dotted line we show for the $K^+/\pi^+$ ratio an estimate
of the effect of higher mass resonances.
The dashed lines show the energy dependence of $T$ and $\mu_b$.}\label{fig_k2pi}
\end{minipage} \end{tabular}
\end{figure}

In Fig.~\ref{fig_tmu} we present the energy dependence of $T$ and $\mu_b$ 
in comparison to results of other similar analyses \cite{cley,nu,zsch,man08}.
In \cite{man08} the strangeness suppression factor $\gamma_s$ is an additional
fit parameter; in \cite{zsch} an inhomogeneous freeze-out scenario is modeled
with spreads in $T$ and $\mu_b$ as extra fit parameters; the approach of 
\cite {letessier05}, employing a full set of non-equilibrium fit parameters,
leads to rather different results compared to those shown here.
We have parametrized our values of $T$ and $\mu_b$ as a function of 
$\sqrt{s_{NN}}$ with the following expressions:
\begin{equation}
T \mathrm{}=T_{lim}\frac{1}{1+\exp(2.60-\ln(\sqrt{s_{NN}(\mathrm{GeV})})/0.45)},
\quad
\mu_b \mathrm{[MeV]}=\frac{1303}{1+0.286\sqrt{s_{NN}(\mathrm{GeV})}}
\label{pt}
\end{equation}
with the "limiting" temperature $T_{lim}$=164 MeV. 

We employ the above parametrizations of $T$ and $\mu_b$ to investigate the 
energy dependence of the production yields of $K^+$ and $\Lambda$ hadrons 
relative to pions, shown in Fig.~\ref{fig_k2pi}.
The $K^+/\pi^+$ ratio shows a rather pronounced maximum 
at a beam energy of 30 AGeV \cite{na49pi}, and the data are well reproduced by 
the model calculations.
In the thermal model this maximum occurs naturally at 
$\sqrt{s_{NN}}\simeq$8 GeV \cite{pbm4}. It is due to the counteracting effects
of the steep rise and saturation of $T$ and the strong monotonous decrease 
in $\mu_b$.
The competing effects are most prominently reflected in the energy dependence
of the $\Lambda$ hyperon to pion ratio (lower panel of Fig.~\ref{fig_k2pi}), 
which shows a pronounced maximum at $\sqrt{s_{NN}}\simeq$5 GeV. 
This is reflected in the $K^+/\pi^+$ ratio somewhat less directly; it appears 
mainly as a consequence of strangeness neutrality, assumed in our calculations.

The model describes the $K^+/\pi^+$ data very well over the full energy range,
as a consequence of the inclusion in the code of the high-mass resonances and 
of the $\sigma$ meson, while our earlier calculations \cite{aa05} were 
overpredicting the SPS data. At RHIC energies, the quality of the present fits 
is essentially unchanged compared to \cite{aa05}, as also the data have changed
somewhat.
The model also describes accurately the $\Lambda/\pi^-$ measurements
as well as those for other hyperons \cite{aat2}.

\begin{figure}[htb]
\vspace{-.6cm}
\begin{tabular}{lr} \begin{minipage}{.56\textwidth}
\hspace{-.4cm}\includegraphics[width=1.07\textwidth]{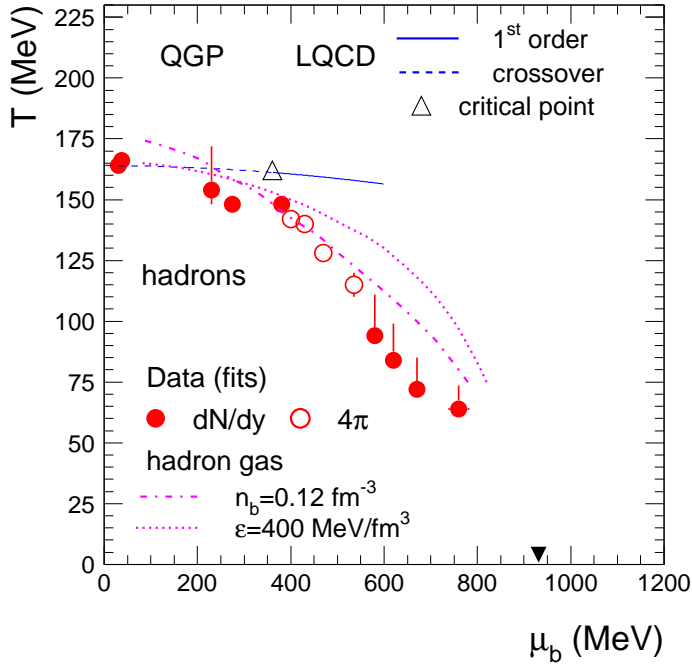}
\end{minipage} &\begin{minipage}{.4\textwidth}
\hspace{-.4cm}\vspace{-1.6cm}
\caption{The phase diagram of strongly interacting matter. The points
represent the results of the thermal fits. For the SPS beam energy of
40 AGeV ($\mu_b\simeq$400 MeV) we show both midrapidity (dN/dy) and full 
phase space (4$\pi$) fit results.
The phase boundary and critical point from lattice QCD (LQCD) calculations
\cite{lqcd} is shown together with freeze-out curves for a hadron gas 
at constant baryon density (baryons and anti-baryons) and energy density.
The full triangle indicates the location of ground state nuclear matter 
(atomic nuclei).
}
\label{t-mu}
\end{minipage} \end{tabular}
\end{figure}

\vspace{-.5cm}
In Fig.~\ref{t-mu} we show the result of our fits in the phase diagram of strongly
interacting matter (for a recent review see \cite{stock09}). 
Our results strongly imply that hadronic observables near and above the horn 
structure at a beam energy of 30-40 AGeV ($\mu_b\simeq$400 MeV), coinciding with
the approach to saturation in $T$, provide a link
to the QCD phase transition. Open questions are whether the chemical
freeze-out curve for larger values of $\mu_b$ actually traces the QCD phase 
boundary or whether chemical freeze-out in this
energy range is influenced by exotic new phases such as have been predicted in
\cite{mclerran_pisarski}. 

In summary, our recent results \cite{aat2} demonstrate that by inclusion of 
the $\sigma$ meson and many higher mass resonances into the resonance spectrum 
employed in the statistical model calculations an improved description is 
obtained of hadron production in central nucleus-nucleus collisions at 
ultra-relativistic energies. A dramatic improvement is visible 
for the $K^+/\pi^+$ ratio, which is now well described at all energies. 
The ``horn'' finds herewith a natural explanation which is, however, 
deeply rooted in and connected to detailed features of the hadronic mass 
spectrum which leads to a limiting temperature and contains the QCD phase 
transition \cite{hagedorn85}. 

\vspace{.2cm}
We acknowledge the support from the Alliance Program of the Helmholtz-Gemeinschaft.

\end{document}